\documentclass[aps,prl,reprint,showpacs]{revtex4-1}

\usepackage{graphicx}
\usepackage{subfigure}
\usepackage{dcolumn}
\usepackage{verbatim}
\usepackage{amsmath}
\usepackage{amssymb}
\usepackage{float}
\usepackage[colorlinks=true,linkcolor=blue,citecolor=blue,filecolor=blue,urlcolor=blue]{hyperref}
\begin{document}

\title{
The SU(3) Algebra in a Cyclic Basis} 

\author{P.~F.~Harrison} 
\email{p.f.harrison@warwick.ac.uk}
\author{R.~Krishnan} \email{k.rama@warwick.ac.uk}
\affiliation{Department of Physics, University of Warwick, Coventry CV4 
7AL, UK}

\author{W.~G.~Scott}
\email{william.scott@stfc.ac.uk}
\affiliation{Rutherford Appleton Laboratory, Chilton, Didcot, OX11
0QX, UK}

\date{\today}

\begin{abstract}
With the couplings between the eight gluons constrained by the structure constants of the su(3) algebra in QCD,
one would expect that there should exist a special basis (or set of bases)  for the algebra wherein, unlike in a Cartan-Weyl basis,
{\em all} \hspace{-0.2mm} gluons interact identically (cyclically) with each other, explicitly on an equal footing.
We report here particular such bases, which we have found in a computer search, and we indicate associated $3 \times 3$ representations.
We conjecture that essentially all cyclic bases for su(3) may be obtained from these making appropriate circulant transformations,
and that cyclic bases may also exist for other su($n$), $n>3$. 
\end{abstract}
\pacs{}

\maketitle

\noindent{\bf 1. Introduction}: Both the su(2) and su(3) algebras
\cite{Note1}
are central to contemporary particle-physics theory,
as the two non-Abelian real Lie algebras generating 
the unitary weak-isospin and colour symmetry groups 
underlying the Standard Model 
of \mbox{(purely-)} weak and strong forces respectively.
The relevant gauge bosons in each case
appear in the adjoint representation of the algebra,
with the self-couplings
among the three weak bosons and among the eight gluons
thereby constrained
by the structure constants of the algebra,
ensuring gauge invariance.

Regarding the weak sector for example \cite{[{}]WEIN,*GLASH,*SALAM}
the su(2) algebra may be readily expressed
by considering basis elements $\hat{w}_i$ satisying the su(2) algebra in the form:
\begin{equation}
 [\hat{w}_1,\hat{w}_2]=\hat{w}_3 \hspace{0.5cm} 
 [\hat{w}_2,\hat{w}_3]=\hat{w}_1 \hspace{0.5cm} 
 [\hat{w}_3,\hat{w}_1]=\hat{w}_2, \hspace{0.5cm} \label{su2c}
\end{equation}
familiar 
as the ordinary 3-space vector-product rule,
or also as the commutation relations of the normalised Pauli matrices 
($\hat{w}_i \leftrightarrow -i\sigma_i/2$, where ${\rm Tr} \, \sigma_i . \sigma_j=2 \delta_{ij}$).
The three corresponding weak-isospin gauge-field components, 
\mbox{$W_1$, $W_2$, $W_3$,} 
may then be incorporated in 
the Lagrangian 
(at least as concerns the pure-gauge part,  see e.g.\ Ref.~\cite{[{}]ATIY,*SMIT})
as coefficients of 
the corresponding 
$3 \times 3$ adjoint representation matrices,
as determined by the structure constants implicit  in Eq.~\ref{su2c}.
The above {\em circulant} or {\em cyclic}  form (Eq.~1) for su(2)  
has the reassuring feature
that clearly all three weak field components will appear on an explicitly equal footing,
with no weak-isospin direction distinguished.

On the other hand, 
the algebra Eq.~\ref{su2c},
when viewed as a complex Lie algebra (namely $A_1$)
expressed in the Chevalley basis, 
takes the canonical (non-cyclic) form: 
\begin{eqnarray}
  [\hat{w}'_3,\hat{w}'_1]=2\hat{w}'_1 \hspace{1.5cm} [\hat{w}'_3,\hat{w}'_2]=-2\hat{w}'_2  \nonumber \\       
 \hspace{0.0cm} [ \hat{w}'_1 , \hat{w}'_2 ] = \hat{w}'_3 \hspace{2.3cm} \label{su2y}
\end{eqnarray}
which, in the present context, might be seen
as better aligned with
the physical states 
after spontaneous symmetry breaking.
In the Standard Model, 
it is the non-zero Higgs field
which picks-out a particular direction in weak-isospin space
to define the 3rd weak-isospin component, 
which then mixes with the U(1) hypercharge boson 
to form the photon and the $Z$-boson,
with the two transverse components 
combined (as the $\pm 1$ eigenstates of $I_3$) 
to describe the charged weak bosons,~$W^+$ and $W^-$.
The two forms
Eq.~\ref{su2c} and Eq.~\ref{su2y} 
(of the complex algebra A$_1$) 
are related by the (complex) transformation: 
\begin{equation}
\left( \begin{matrix} \hat{w}'_1 \cr  \hat{w}'_2 \cr   \hat{w}'_3  \end{matrix} \right) \! \!   =   \! \!
\left( \begin{matrix} i \:  & \: -1 &   \cr 
                i \: & \: +1 &    \cr  
                  &  &  \: 2i \end{matrix}  \right) \! \!
\left( \begin{matrix} \hat{w}_1 \cr  \hat{w}_2 \cr   \hat{w}_3  \end{matrix} \right). \! \!  
\label{su2t}
\end{equation}
While Eq.~\ref{su2y} may be viewed as an expression of 
the (real but non-compact) Lie algebra 
sl(2,R) $\simeq$ su(1,1), 
the corresponding representation matrices
can 
still be utilised in constructing the Lagrangian,
provided that they are coupled with
the appropriately defined (complex) field components
(proportional to
\mbox{$W^{+}=(W_1-iW_2)/\sqrt{2}$},
\mbox{$W^{-}=(W_1+iW_2)/\sqrt{2}$} and $W_3$)
such that the action remains unchanged. 
Naturally, the theory remains SU(2) gauge invariant,
and all physical consequences,
before or after spontaneous symmetry breaking, 
remain undisturbed by such a change of basis.

In the case of the strong interaction (QCD)
the SU(3) colour symmetry is well-known to be unbroken,
and we may expect that 
the theoretical prediction of any measurable
will always result as summed or averaged symmetrically over colours 
and hence be readily expressible in an explicitly basis-independent way.
Somewhat analogous to the Pauli matrices for su(2),
in the case of su(3) one has the Gell-Mann matrices, $\lambda_a$~\cite{[{}]GELL,*NEEM}, $a=1 \dots 8$
(with ${\rm Tr} \, \lambda_a . \lambda_b =2 \delta_{ab}$, $a,b=1 \dots 8$),     
which constitute a $ 3 \times 3$ matrix representation 
of the su(3) algebra in the form ($\hat{g}_a \leftrightarrow -i\lambda_a/2$): 
\begin{eqnarray}
[\hat{g}_1,\hat{g}_2]=\hat{g}_3 \hspace{8mm} 
[\hat{g}_1,\hat{g}_4]=\hat{g}_7/2 \hspace{8mm} [\hat{g}_1,\hat{g}_5]=-\hat{g}_6/2  \hspace{9mm}  \nonumber \\
\hspace{1mm} [\hat{g}_2,\hat{g}_4]=\hat{g}_6/2 \hspace{3mm} [\hat{g}_2,\hat{g}_5]=\hat{g}_7/2 \hspace{3mm}
 [\hat{g}_4,\hat{g}_5]=\hat{g}_3/2+\sqrt{3}/2 \hat{g}_8  \hspace{3mm} \nonumber \\
\hspace{1mm} [\hat{g}_4,\hat{g}_8]=-\sqrt{3}/2 \hat{g}_5  \hspace{9mm} 
 [\hat{g}_6,\hat{g}_7]=-\hat{g}_3/2+\sqrt{3}/2 \hat{g}_8 \hspace{11mm}  \label{su3g} 
\end{eqnarray}
where the Lie brackets quoted (Eq.~\ref{su3g}) are sufficient
to determine all non-zero brackets, 
given that the su(3) structure constants 
are totally antisymmetric in the Gell-Mann basis.
One sees almost immediately however that the su(3) algebra
in the form specified by Eq.~\ref{su3g} is far from cyclic (cf.\ Eq.~\ref{su2c}),
whereby the Pauli basis for su(2) and Gell-Mann basis for su(3) 
cannot be considered fully analogous, at least in this respect.

Choice of basis for the algebra being of conceptual importance here,
we find ourselves 
motivated to ask 
if a cyclic basis, 
with all eight gluons in the same
relative relationship,
analogous to Eq.~1 for su(2), 
actually exists (or not) for the case of su(3).
Our best literature search 
uncovered just one allusion \cite{ROSSI} 
to a ``cyclic basis'' (Eq.~\ref{su2c}) for su(2),
but no mention of any such equivalent for su(3),
or indeed for any su($n$), $n > 2$ (or for any other Lie algebra). 
Having thus had 
to rely on our own efforts in attempting to answser this question,
we are now able to report
what we believe to be 
the complete set of such cyclic bases for su(3)
(and for a few other very closely related Lie algebras).
In the following sections 
we briefly outline our methods
and proceed to document the cyclic forms we obtained.  
As a by-product, 
we are able to give 
a set of $3 \times 3$ matrices 
which constitute a matrix representation 
of the su(3) algebra in cyclic form,
very closely analogous to the Pauli matrices for su(2).
These matrices 
(at least as regards their general form/symmetries etc.) 
in fact turn out
to be familiar to us already 
in a somewhat different 
particle-physics context \cite{HS94,*SCOTT}, as will be detailed later (see Section~4).  \\

\noindent {\bf 2. A `Theorem' and a Computer Search:}
\noindent In a heuristic spirit,
let us suppose
that such cyclic forms exist 
for at least some Lie algebras beyond su(2).
In particular, taking su(3) as an example,
we would then expect
(in analogy to Eq.~\ref{su2c} for the su(2) case) 
to be able to write all brackets in the form:
\begin{eqnarray}
[\hat{g}_a,\hat{g}_b] & = & f_{ab}^c \, \hat{g}_c \hspace{0.4cm} {\rm with} \hspace{0.3cm} 
f_{a+x \, b+x}^{c+x}=f_{ab}^c,  \hspace{0.4cm} \nonumber  \\
 {\rm where} \hspace{0.1cm} & &  x = 1\dots8 \hspace{0.8cm}
{\rm indices} \hspace{0.2cm} {\rm mod} \hspace{0.2cm} 8,1 \hspace{0.8cm} \label{genc}
\end{eqnarray}
and where ``indices mod 8,1'' implies
a cyclic interpretation of subscripts and superscipts, such that $7+1=8$, $8+1=1$ etc.
The algebra being expressed in this form,
with all basis elements, $\hat{g}_1 \dots \hat{g}_8$, appearing 
cyclically on an explicitly equal footing,
precisely generalises the su(2) case (Eq.~\ref{su2c}).
We add that
it will prove useful 
to visualise the base elements in such a basis
\begin{figure}[h]
\begin{center}
\includegraphics[scale=0.43]{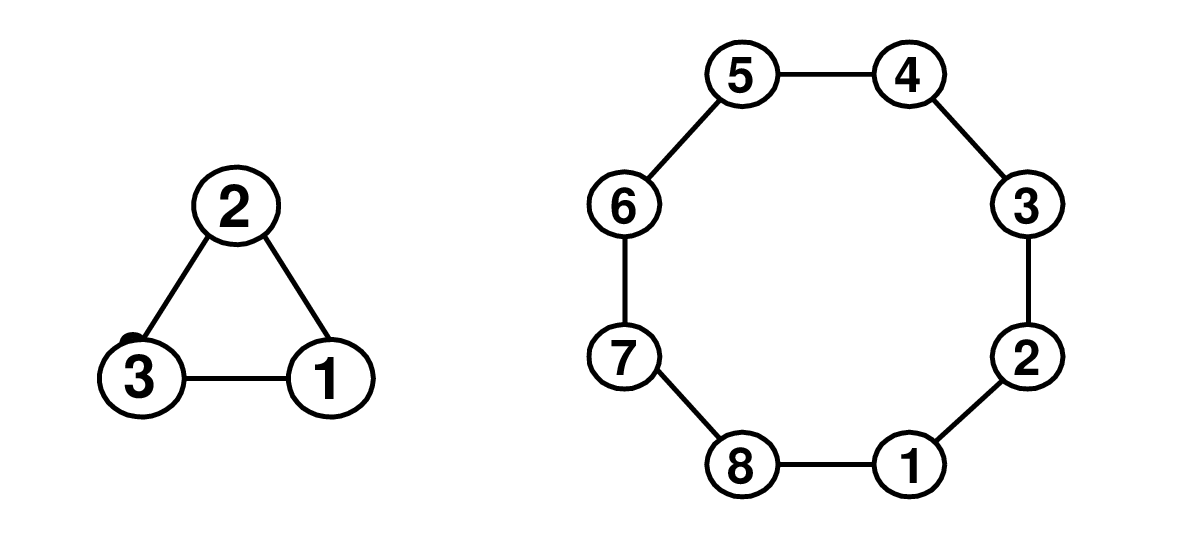}
\caption{ 
In a cyclic basis,   
the base elements of su(2) and su(3)
are usefully visualised as being located at the vertices 
of an equilateral triangle and regular octagon respectively.}
\label{fig:trioct}
\end{center}
\end{figure}
as being located at the vertices of a regular polygon, 
i.e.\ at the corners of an equilateral triangle in the su(2) case,
or at the corners of a regular octagon in the case of su(3),
as illustrated (for subsequent reference) in Figure~1.

Now, given a real Lie algebra such as su(3),
we can always transform to a new basis 
with any real non-singular linear transformation we choose.
Along with the transformation of the basis elements, $\hat{g}_a$, 
the structure constants, $f_{ab}^c$, will transform, 
in a tensor-like fashion, 
with two covariant and one contravariant index.
Applying such a transformation 
(with transformation matrix $C$) to the cyclic form, Eq.~\ref{genc}, above:
\begin{equation}
(\hat{g}_a')^T \, \, =  \, \, C \, . \, (\hat{g}_a)^T \label{tranc}
\end{equation}
where $(\hat{g}_a)=(\hat{g}_1,\dots,\hat{g}_8)$ and 
$(\hat{g}_a')=(\hat{g}_1',\dots,\hat{g}_8')$, 
we take it as self-evident that, 
{\em if} 
the transformation 
takes the form of a (+1)-circulant matrix \cite{DAVIS}:
\begin{equation}
C= {\rm circ}(c_1,c_2,c_3,c_4,c_5,c_6,c_7,c_8) \label{circ}
\end{equation}
that is, if:
\begin{equation}
\left( \begin{matrix} \hat{g}_1' \cr \hat{g}_2' \cr \hat{g}_3' \cr \hat{g}_4'
\cr \hat{g}_5' \cr \hat{g}_6' \cr \hat{g}_7' \cr \hat{g}_8' \end{matrix} \right)
=
\left( \begin{matrix} 
c_1 & c_2 & c_3 & c_4 & c_5 & c_6 & c_7 & c_8 \cr
c_8 & c_1 & c_2 & c_3 & c_4 & c_5 & c_6 & c_7 \cr
c_7 & c_8 & c_1 & c_2 & c_3 & c_4 & c_5 & c_6 \cr
c_6 & c_7 & c_8 & c_1 & c_2 & c_3 & c_4 & c_5 \cr
c_5 & c_6 & c_7 & c_8 & c_1 & c_2 & c_3 & c_4 \cr
c_4 & c_5 & c_6 & c_7 & c_8 & c_1 & c_2 & c_3 \cr
c_3 & c_4 & c_5 & c_6 & c_7 & c_8 & c_1 & c_2 \cr
c_2 & c_3 & c_4 & c_5 & c_6 & c_7 & c_8 & c_1 \cr \end{matrix} \right)
\left( \begin{matrix} \hat{g}_1 \cr \hat{g}_2 \cr \hat{g}_3 \cr \hat{g}_4 
\cr \hat{g}_5 \cr \hat{g}_6 \cr \hat{g}_7 \cr \hat{g}_8 \end{matrix} \right),   \label{trancirc}
\end{equation}
{\em then} the transformed algebra (basis $\hat{g}_a'$) will also be cyclic.
This `theorem' clearly generalises to complex transformations,
except of course that complex transformations
may result in a different real form of the same (complex) algebra.
This suggests 
that if we were just able to discover one cyclic form for su(3),
we could immediately generate essentially all possible cyclic forms
(for su(3) itself -  or indeed for its complex extension A$_2$ and hence also for the sl(3,R) or su(2,1) real forms)
by making appropriate (real or complex) circulant transformations.

At this point therefore, 
we resorted to a computer-based search
aimed at finding at least one cyclic expression
for the su(3) algebra.
In a cyclic basis, 
in the 8-dimensional case (Figure~1 right), the four brackets:
\begin{eqnarray}
 [ \hat{g}_1,\hat{g}_2 ]
& = & f_{12}^1 \, g_1+f_{12}^2 \, g_2+f_{12}^3 \, g_3+f_{12}^4 \, g_4 \nonumber \\
& & + f_{12}^5 \, g_5+f_{12}^6 \, g_6+f_{12}^7 \, g_7+f_{12}^8 \, g_8\hspace{0.6cm} \label{su12} \\
 \hspace{0.0cm} [ \hat{g}_1,\hat{g}_3 ] 
& = & f_{13}^1 \, g_1+f_{13}^2 \, g_2+f_{13}^3 \, g_3+f_{13}^4 \, g_4 \nonumber \\
& & + f_{13}^5 \, g_5+f_{13}^6 \, g_6+f_{13}^7 \, g_7+f_{13}^8 \, g_8\hspace{0.6cm} \label{su13} \\
 \hspace{0.0cm} [ \hat{g}_1,\hat{g}_4 ] 
& = & f_{14}^1 \, g_1+f_{14}^2 \, g_2+f_{14}^3 \, g_3+f_{14}^4 \, g_4 \nonumber \\
& & + f_{14}^5 \, g_5+f_{14}^6 \, g_6+f_{14}^7 \, g_7+f_{14}^8 \, g_8\hspace{0.6cm} \label{su14} \\ 
 \hspace{0.0cm} [ \hat{g}_1,\hat{g}_5 ] 
& = & f_{15}^1 \, g_1+f_{15}^2 \, g_2+f_{15}^3 \, g_3+f_{15}^4 \, g_4 \nonumber \\
& & + f_{15}^5 \, g_5+f_{15}^6 \, g_6+f_{15}^7 \, g_7+f_{15}^8 \, g_8, \hspace{0.5cm}  \label{su15}
\end{eqnarray}
together with the cyclic constraint (Eq.~\ref{genc})  
and the Lie antisymmetry condition,
are clearly sufficient to readily determine all brackets. 
Indeed, taking 
the cyclic constraint 
and the Lie antisymmetry condition together, Eq.~\ref{su15} 
is readily re-cast to involve just  four parameters as follows:
\begin{eqnarray}
 \hspace{0.0cm} [ \hat{g}_1,\hat{g}_5 ] 
& = & f_{15}^1 \, g_1+f_{15}^2 \, g_2+f_{15}^3 \, g_3+f_{15}^4 \, g_4 \nonumber \\
& & - f_{15}^1 \, g_5-f_{15}^2 \, g_6-f_{15}^3 \, g_7-f_{15}^4 \, g_8 .   \hspace{0.6cm} \label{su15m}
\end{eqnarray}
For a manageable computer search,
in place of Eq.~\ref{su15m} 
(and thereby risking to miss valid instances), 
we in fact implemented the simpler 
condition:
\begin{eqnarray}
 \hspace{0.0cm} [ \hat{g}_1,\hat{g}_5 ] & = & 0. \hspace{0.4cm}  \label{su15f}
\end{eqnarray} 
Even so, we still had 24 structure constants
$f_{12}^c$, $f_{13}^c$, $f_{14}^c$, $c=1 \dots 8$ 
(Eqs.~\ref{su12}-\ref{su14}) to find,
and our search was therefore restricted 
to trying only the values $+1$, $-1$ and $0$ 
for each of these 24 parameters. 
After $\sim \! \! 100$ hours running on the RAL PPD linux farm \cite{BREW}, 
we found that, out of $3^{24} \simeq 2.8 \times 10^{11}$ possibilities, 
a total of 972 choices 
fully satisfied the Jacobi relation
(excluding cases where any of $f_{12}^c$, $f_{13}^c$, $f_{14}^c$=0, $\forall c=1 \dots 8$).
Then, in a second pass, 
selecting only      
cases with negative-definite Killing form
(corresponding to the real algebra su(3) itself),
just two cases remained as displayed in Table~1,
\begin{table}[h]
\begin{tabular}{|c|c|c|c|c|} \hline
Case 
& $f_{12}^c$ & $f_{13}^c$ & $f_{14}^c$ & metric \rule[0mm]{0mm}{7mm} \\ 
No.\  
 & $1 \, 2 \, 3 \, 4 \, 5 \, 6 \, 7 \,  8$ 
                        & $1 \, 2 \, 3 \, 4 \,  5 \,  6 \,  7 \,  8$ 
                                                & $1 \, 2 \, 3 \, 4 5 \, 6 \, 7 \, 8$ & signature \rule[-3mm]{0mm}{3mm}  \\  \hline
 1 \rule[-1mm]{0mm}{7mm} 
 & 
$0 \, 0 \, 1 \, 1 \,  0 \,  0 \, \bar{1} \, 1$ & 
$0 \, \bar{1} \, 0 \, \bar{1} \, 0  \, \bar{1} \,  0 \, 1$ & 
$0 \, \bar{1} \, 1 \, 0 \, 0 \, \bar{1} \, \bar{1} \, 0$      & $(-24)^8$ \\
 2 \rule[-1mm]{0mm}{7mm} 
& 
$0 \, 0 \, 0 \, 1 \, 0 \, 0 \, \bar{1} \, 1$ & 
$0 \, \bar{1} \,  0 \, \bar{1} \, 0 \, \bar{1} \, 0 \, 0$ & 
$0 \, \bar{1} \, 1 \, 0 \, 0 \, \bar{1} \,  0 \, 0$       & $(-18)^4$, $(-6)^4$ \\ \hline
\end{tabular}
\caption{
The two sets 
of cyclic structure constants 
for the Lie algebra su(3) 
found in our computer search
(only structure-constant values 
of $+1$, $0$, $-1$ were tried, 
as indicated here by $1$, $0$, $\bar{1}$ respectively).
The last column gives the eigenvalues
of the Killing metric 
with their multiplicities.}
\end{table} 
excluding cases trivially related 
to these by overall sign change, or/and by
simple reversal of the cyclic ordering - see below.


\noindent {\bf 3. Transformations between Cyclic Forms:}
Directly from Table~1: \hspace{-2mm} case~1 \hspace{0.1mm} then,
we have that the su(3) algebra 
may be expressed 
in the cyclic form:
\begin{eqnarray}
\left[\hat{g}^{(1)}_{a+1},\hat{g}^{(1)}_{a+2}\right] & = & \hspace{2.5mm}
\hat{g}^{(1)}_{a+3}+\hat{g}^{(1)}_{a+4}-\hat{g}^{(1)}_{a+7}+\hat{g}^{(1)}_{a+8} \label{su12c2} \\
\left[\hat{g}^{(1)}_{a+1},\hat{g}^{(1)}_{a+3}\right] & = & 
-\hat{g}^{(1)}_{a+2}-\hat{g}^{(1)}_{a+4}-\hat{g}^{(1)}_{a+6}+\hat{g}^{(1)}_{a+8} 
\label{su13c2} \\
\left[\hat{g}^{(1)}_{a+1},\hat{g}^{(1)}_{a+4}\right] & = & 
-\hat{g}^{(1)}_{a+2}+\hat{g}^{(1)}_{a+3}-\hat{g}^{(1)}_{a+6}-\hat{g}^{(1)}_{a+7} \label{su14c2} \\
\left[ \hat{g}^{(1)}_{a+1},\hat{g}^{(1)}_{a+5} \right]  & = &  0, \hspace{0.4cm} 
\hspace{0.3cm} a=1 \dots 8  \hspace{0.5cm} {\rm mod} \; 8,1 .  \hspace{1.0cm} \label{su15c2}
\end{eqnarray}
The Killing form is diagonal
in this basis 
($\kappa^{(1)}_{a b}
=-24\delta_{ab}$) 
and the structure constants are totally antisymmetric.
As regards economy of expression, 
it turns out that all structure constants Eqs.~\ref{su12c2}-\ref{su15c2}
could if necessary be readily inferred from Eq.~\ref{su13c2} alone,
exploiting the total antisymmetry and the cyclic property (Eq.~\ref{genc}) together.

From Table~1: \hspace{-3mm} case~2, we have that
the real algebra su(3) may also be expressed: 
\begin{eqnarray}
\left[\hat{g}^{(2)}_{a+1},\hat{g}^{(2)}_{a+2}\right] & = & \hspace{2.5mm}
                       \hat{g}^{(2)}_{a+4}-\hat{g}^{(2)}_{a+7}+\hat{g}^{(2)}_{a+8} \label{su12c3} \\
\left[\hat{g}^{(2)}_{a+1},\hat{g}^{(2)}_{a+3}\right] & = &
                       -\hat{g}^{(2)}_{a+2}-\hat{g}^{(2)}_{a+4}-\hat{g}^{(2)}_{a+6}  
\label{su13c3} \\
\left[\hat{g}^{(2)}_{a+1},\hat{g}^{(2)}_{a+4}\right] & = & 
                     -\hat{g}^{(2)}_{a+2}+\hat{g}^{(2)}_{a+3}-\hat{g}^{(2)}_{a+6} \label{su14c3} \\
\left[ \hat{g}^{(2)}_{a+1},\hat{g}^{(2)}_{a+5} \right]  & = &  0, \hspace{0.4cm} 
\hspace{0.3cm} a=1-8  \hspace{0.5cm} {\rm mod} \; 8,1 .  \hspace{1.0cm} \label{su15c3}
\end{eqnarray}
In this case the Killing form is circulant but not diagonal:
\begin{equation}
\kappa^{(2)}
={\rm circ}\{-12,0,0,0,6,0,0,0\} .
\end{equation}
The transformation from su(3) (Eqs.~\ref{su12c2}-\ref{su15c2}) to su(3) (Eqs.~\ref{su12c3}-\ref{su15c3}) is given by:
\begin{equation}
C^{(21)}={\rm circ}\{(1+\sqrt{3})/4,0,0,0,(1-\sqrt{3})/4,0,0,0\}
\end{equation}
and the corresponding inverse transformation by:
\begin{equation}
C^{(12)}=
{\rm circ}\{(1+1/\sqrt{3}),0,0,0,(1-1/\sqrt{3}),0,0,0\}.
\end{equation}

Cyclic forms satisfying Eq.~\ref{genc}
but related to case~1 and case~2
by trivial reversal of cyclic ordering and so excluded from Table~1,
may then be generated by 
a \mbox{(-1)-circulant} (or retro-circulant \cite{DAVIS}) transformation:
\begin{equation}
C^{(\bar{i}i)}=
(-1){\rm circ}\{(0,0,0,0,0,0,0,1\} 
\label{retro}
\end{equation}
having non-zero (unit) entries only  on the trailing diagional.
Our claim to have found the complete set of cyclic bases for su(3) 
and hence for the complex algebra $A_2$ and any of its real forms,
rests on the plausible conjecture that any such basis may be reached 
with a combination of the transformation Eq.~\ref{retro} 
and general circulant transformations Eqs.~\ref{tranc}-\ref{trancirc},
with arbirary complex parameters. 

\noindent {\bf 4. Relation to the Gell-Mann Basis and the Gell-Mann Matrices:}
The Gell-Mann basis of su(3) is defined by Eq.~\ref{su3g}. 
If we define:
\begin{equation}
x_-=1/\sqrt{3}-1 \hspace{0.4cm} x_0=-2/\sqrt{3} \hspace{0.4cm} x_+=1/\sqrt{3}+1,
\end{equation}

\noindent then the (real) transformation from cyclic su(3) 
(taking as an example the particular cyclic form Eqs.~\ref{su12c2}-\ref{su15c2}) 
to the Gell-Mann basis (Eq.~\ref{su3g}) is given by:
\begin{equation}
\left( \begin{matrix}
 0 & -\frac{x_+}{8} & \frac{x_-}{8} & -\frac{x_0}{8} &
           0 & -\frac{x_-}{8} & \frac{x_+}{8} & -\frac{x_0}{8} \cr
 0 &  \frac{x_-}{8} & -\frac{x_+}{8} & -\frac{x_0}{8} &
           0 & -\frac{x_+}{8} & \frac{x_-}{8} & \frac{x_0}{8} \cr
-\frac{\sqrt{3} \, x_-}{8} & 0 & 0 & 0 & 
                        \frac{\sqrt{3} \, x_+}{8} & 0 & 0 & 0 \cr
  0 & -\frac{x_0}{8} & \frac{x_0}{8} & -\frac{x_0}{8} &
           0 & -\frac{x_0}{8} & \frac{x_0}{8} & -\frac{x_0}{8} \cr 
  0 & -\frac{x_0}{8} & \frac{x_0}{8} & \frac{x_0}{8} &
           0 & \frac{x_0}{8} & -\frac{x_0}{8} & -\frac{x_0}{8} \cr
 0 & -\frac{x_-}{8} & \frac{x_+}{8} & -\frac{x_0}{8} &
           0 & -\frac{x_+}{8} & \frac{x_-}{8} & -\frac{x_0}{8} \cr
 0 &  \frac{x_+}{8} & -\frac{x_-}{8} & -\frac{x_0}{8} &
           0 & -\frac{x_-}{8} & \frac{x_+}{8} & \frac{x_0}{8} \cr
-\frac{\sqrt{3} \, x_+}{8} & 0 & 0 & 0 &
                          -\frac{\sqrt{3} \, x_-}{8} & 0 & 0 & 0
\end{matrix} \right).
\end{equation}
The corresponding inverse transformation from the Gell-Mann basis (Eq.~\ref{su3g}) 
back to the cyclic basis (Eqs.~\ref{su12c2}-\ref{su15c2}) is given by the coresponding inverse matrix:
\begin{equation}
\left( \begin{matrix}
 0 & 0 & -\sqrt{3} \, x_- & 0 &
           0 & 0 & 0 & -\sqrt{3} \, x_+ \cr
 -x_+ & x_- & 0 & -x_0 &
           -x_0 & -x_- & x_+ & 0 \cr
x_-  & -x_+ & 0 & x_0 & 
             x_0  & x_+ & -x_- & 0 \cr
  -x_0 & -x_0 & 0 & -x_0 &
           x_0 & -x_0 & -x_0 & 0 \cr 
  0 & 0 & \sqrt{3} \, x_+ & 0 &
                0 & 0 & 0 & -\sqrt{3} \, x_- \cr
 -x_- & -x_+ & 0 & -x_0 &
              x_0 & -x_+ & -x_- & 0 \cr
 x_+ &  x_- & 0 & x_0 &
              -x_0 & x_- & x_+ & 0 \cr
 -x_0 & x_0 & 0 & -x_0 &
                          -x_0  & -x_0 & x_0 & 0
\end{matrix} \right) .
\end{equation}


Starting from the Gell-Mann matrices
and exploiting the above transformation,
we 
may now readily
construct a $3 \times 3$ (i.e. the fundamental) representation 
of su(3) in cyclic form (again, by way of example, in the particular form Eqs.~\ref{su12c2}-\ref{su15c2}).
Further defining:
\begin{equation}
b=\frac{x_-}{3} \omega + \frac{x_0}{3} + \frac{x_+}{3} \bar{\omega}     \hspace{0.8cm}
\bar{b}= \frac{x_-}{3} \bar{\omega} + \frac{x_0}{3} + \frac{x_+}{3} \omega
\end{equation} 
and normalising with 
normalisation constant 
$N=2$,
we find that the $3 \times 3$ matrices:
\begin{eqnarray}
\lambda_1^{(1)}= N 
                \left( \begin{matrix}  x_- & 0 & 0 \cr 
                              0 & x_0 & 0 \cr
                              0 & 0 & x_+   \end{matrix} \right)                      & \hspace{0.2cm} & 
\lambda_2^{(1)} = -N 
          \left( \begin{matrix} 0 & b \omega & \bar{b} \cr
         \bar{b} \bar{\omega} & 0 & b \bar{\omega} \cr
            b & \bar{b} \omega &  0  \end{matrix} \right) \nonumber \\
\lambda_3^{(1)} = N 
           \left( \begin{matrix}  0 & b \bar{\omega} & \bar{b} \cr
               \bar{b} \omega & 0 & b \omega \cr
               b &  \bar{b} \bar{\omega} & 0   \end{matrix} \right)    & \hspace{0.2cm}  &
\lambda_4^{(1)} = -N 
            \left( \begin{matrix} 0 & \bar{b} & b \cr
                            b & 0 & \bar{b} \cr
                           \bar{b} & b & 0   \end{matrix} \right) \nonumber \\
\lambda_5^{(1)} = N 
             \left( \begin{matrix}  x_+ & 0 & 0 \cr
                            0 &  x_0 & 0 \cr
                            0 & 0 &  x_-   \end{matrix} \right)                       & \hspace{0.2cm} &
\lambda_6^{(1)} = -N 
            \left( \begin{matrix}   0 & \bar{b} \omega & b \cr
             b \bar{\omega} & 0 & \bar{b} \bar{\omega} \cr
               \bar{b} & b \omega & 0   \end{matrix} \right) \nonumber \\
\lambda_7^{(1)} = N 
               \left( \begin{matrix} 0 & \bar{b} \bar{\omega} & b \cr
                b \omega & 0 & \bar{b} \omega \cr
                \bar{b} & b \bar{\omega} & 0 \end{matrix} \right)  & \hspace{0.2cm} &
\lambda_8^{(1)} = -N 
                   \left( \begin{matrix} 0 & b & \bar{b} \cr
                         \bar{b} & 0 & b \cr
                         b & \bar{b} & 0   \end{matrix} \right) \label{hrs2}
 \end{eqnarray} 

\noindent
do indeed constitute a matrix representation 
($\hat{g}_a \leftrightarrow -i\lambda_a/2$) of su(3) 
in the cyclic form Eqs.~\ref{su12c2}-\ref{su15c2}. 
Normalising Eq.~\ref{hrs2} rather with 
$N=1/\sqrt{2}$,
we reproduce the condition ${\rm Tr} \, \lambda_a . \lambda_b =2 \delta_{ab}$, $a,b=1 \dots 8$,
and the cyclic su(3) algebra Eqs.~\ref{su12c2}-\ref{su15c2} becomes instead:
\begin{eqnarray}
\left[\hat{g}^{(1)}_{a+1},\hat{g}^{(1)}_{a+2}\right] & = & \hspace{2.5mm}
\frac{1}{2\sqrt{2}}(\hat{g}^{(1)}_{a+3}+\hat{g}^{(1)}_{a+4}-\hat{g}^{(1)}_{a+7}+\hat{g}^{(1)}_{a+8})  \hspace{4mm} \\ 
\left[\hat{g}^{(1)}_{a+1},\hat{g}^{(1)}_{a+3}\right] & = & 
\frac{1}{2\sqrt{2}}(-\hat{g}^{(1)}_{a+2}-\hat{g}^{(1)}_{a+4}-\hat{g}^{(1)}_{a+6}+\hat{g}^{(1)}_{a+8}) \hspace{4mm}  \\
\left[\hat{g}^{(1)}_{a+1},\hat{g}^{(1)}_{a+4}\right] & = & 
\frac{1}{2\sqrt{2}}(-\hat{g}^{(1)}_{a+2}+\hat{g}^{(1)}_{a+3}-\hat{g}^{(1)}_{a+6}-\hat{g}^{(1)}_{a+7})  \hspace{4mm} \\ 
\left[ \hat{g}^{(1)}_{a+1},\hat{g}^{(1)}_{a+5} \right]  & = &  0, \hspace{0.4cm} 
 \hspace{0.4cm} a=1 \dots 8  \hspace{0.5cm} {\rm mod} \; 8,1 .  \hspace{1.0cm}  
\end{eqnarray}
These cyclic structure constants could now 
be used directly in the QCD Lagrangian, 
in place of the Gell-Mann structure constants in the pure-gauge part,
with the matrices Eq.~\ref{hrs2} 
(with $N=1/\sqrt{2}$) 
replacing the Gell-Mann matrices as concerns the interaction with the quarks.

Note that from Eq.~\ref{su15f}, `opposite' pairs of generators 
(i.e.\ diametrically `opposite' in Figure~1) commute,
whereby the two corresponding `opposite' matrices 
(in Eqs.~\ref{hrs2})    
have always very similar form.
In particular $\lambda_1$ and $\lambda_5$ both turn out to be diagonal here.
The remaining (off-diagonal) matrices, 
$\lambda_2,\lambda_3,\lambda_4,\lambda_6, \lambda_7,\lambda_8$,
exhibit a characteristic $0^o$ or $\pm120^o$ phase drop 
between cyclically-related off-diagonal elements equal in modulus, 
and thereby have the symmetry of the circulant/circulative\cite{CHEN} 
forms introduced in Ref.~\cite{HS94}.
Indeed,
simply adding a component proportional to the $3 \times 3$ identity 
to each generator $\lambda_a$, 
Eqs.~\ref{hrs2},     
(or even just exponentiating the $\lambda_a$) 
produces precisely the matrix forms discussed in Ref.~\cite{HS94,*SCOTT} which, 
taken to act in the generation space as candidate fermion mass matrices,
led us in 1994 
to the prediction of `trimaximal' mixing for quarks at very high energies~\cite{HS94}. 
(In fact matrices somewhat similar to Eq.~\ref{hrs2} 
have been proposed previously \cite{[{}]PATER,*FAIR1,*FAIR2,*FAIR3} in the su(3) context, 
but not yielding explicit cyclic forms for the su(3) algebra as in the present work.) 
It may be readily verified 
(diagonalising the individual $\lambda_a$, Eqs.~\ref{hrs2}),    
that the relative transformation
between any two `non-opposite' $\lambda_a$ (Figure~1) 
always takes the form of a trimaximal (i.e.\ $3 \times 3$ `flat'\cite{FLAT} unitary) matrix,
thereby to some degree confirming the underlying intuitive notion 
that our cyclic generators and their corresponding representation matrices
are, in some meaningful sense, equally distributed (and maximally separated) 
in the space in which they act. \\

\noindent {\bf 5. Summary and a Final Conjecture:} 
We have presented 
cyclic expressions 
for the real Lie algebra su(3),
relevant in particle physics as the gauge group of QCD,
such that all gluons are seen to interact on an explicitly equal footing. 
We have plausibly conjectured that all cyclic forms 
of the complex algebra A$_2$ (and its real forms)
may be readily generated from the forms given here,
using appropriate circulant transformations.
We have given $3 \times 3$ representations
of these cyclic forms for su(3) which faithfully generalise the $2 \times 2$ Pauli matrices. 

We close with a final (tentative) conjecture
that cyclic forms (analogous to Eq.~\ref{genc}) should exist 
for other Lie algebras, at the very least for su($n$), $n>3$. 
This supposition is based on the physical notion that in a Grand Unified Theory
(in which strong, weak and electromagnetic forces unify within a single gauge group, e.g.\ SU(5) \cite{GUTS}),
nothing should {\it a priori} distinguish 
the various gauge bosons in the unbroken theory. 
Whether cyclic bases have any computational advantages 
in practical calculations
(e.g.\ in the search for classical solutions, in lattice calculations  etc.) remains to be seen. 
\\

\noindent {\bf Acknowledgement:}
PFH acknowledges support from the UK Science and Technology Facilities Council (STFC)
on the STFC consolidated grant ST/H00369X/1.
PFH and RK also acknowledge the hospitality of the Centre for Fundamental
Physics (CfFP) at the Rutherford Appleton Laboratory (RAL). RK acknowledges support from
CfFP and the University of Warwick. 
WGS thanks Chan Hong Mo (RAL) for helpful discussions. \\

\noindent {\bf Appendix:} 

\noindent
For completeness, it should be said that 
in our computer search, 
if instead of requiring negative-definite Killing form in the second pass,
we require only that the Killing form have non-zero determinant 
(i.e. if we require only that the algebra be semi-simple, $ | \kappa | \ne 0$)
then, 
in addition to case~1 and case~2 
from the main text above 
(reproduced once again in Table~II below),
we find also the especially  straightforward 
cyclic form Table~II case~3.

\begin{table}[h]
\begin{tabular}{|c|c|c|c|c|} \hline
Case 
& $f_{12}^c$ & $f_{13}^c$ & $f_{14}^c$ & metric \rule[0mm]{0mm}{7mm} \\ 
No.\  
 & $1 \, 2 \, 3 \, 4 \, 5 \, 6 \, 7 \,  8$ 
                        & $1 \, 2 \, 3 \, 4 \,  5 \,  6 \,  7 \,  8$ 
                                                & $1 \, 2 \, 3 \, 4 5 \, 6 \, 7 \, 8$ & signature \rule[-3mm]{0mm}{3mm}  \\  \hline
 1 \rule[-1mm]{0mm}{7mm} 
 & 
$0 \, 0 \, 1 \, 1 \,  0 \,  0 \, \bar{1} \, 1$ & 
$0 \, \bar{1} \, 0 \, \bar{1} \, 0  \, \bar{1} \,  0 \, 1$ & 
$0 \, \bar{1} \, 1 \, 0 \, 0 \, \bar{1} \, \bar{1} \, 0$      & $(-24)^8$ \\
 2 \rule[-1mm]{0mm}{7mm} 
& 
$0 \, 0 \, 0 \, 1 \, 0 \, 0 \, \bar{1} \, 1$ & 
$0 \, \bar{1} \,  0 \, \bar{1} \, 0 \, \bar{1} \, 0 \, 0$ & 
$0 \, \bar{1} \, 1 \, 0 \, 0 \, \bar{1} \,  0 \, 0$       & $(-18)^4$, $(-6)^4$ \\ 
 3 \rule[-1mm]{0mm}{7mm} 
 & 
$0 \, 0 \, 1 \, 0 \,  0 \,  0 \,  0 \,  0$ & 
$0 \, 0 \, 0 \, 0\,  0 \,  0 \,  0 1$ & 
$0 \, 0 \, 0 \, 0 \, 0 \, 0 \, \bar{1} \, 0$      & $(-6)^4$, $(+6)^4$ \\ 
\hline
\end{tabular}
\caption{As for Table~I, except that
instaed of requiring negative definite Killing form,
we require only that the Killing form have non-zero determinant 
(i.e.\ we require only that the algebra be semi-simple, $ | \kappa | \ne 0$).
While case~1 and case~2 
are reproduced excatly as in  Table~I,
the additional 
form case~3 (c.f.\ Table~I) 
 corresponds to the non-compact algebra su(2,1).}
\end{table} 
\noindent 
With metric signature $(-^4,+^4)$, %
Table~II~case~3 
evidently gives a cyclic form of the non-compact algebra su(2,1): 
\begin{eqnarray}
\left[ \hat{g}^{(3)}_{a+1},\hat{g}^{(3)}_{a+2} \right]  & = &  \hat{g}^{(3)}_{a+3} \hspace{0.4cm} \label{sl12} \\
\left[ \hat{g}^{(3)}_{a+1},\hat{g}^{(3)}_{a+3} \right]  & = &  \hat{g}^{(3)}_{a+8} \hspace{0.4cm} 
\label{sl13} \\
\left[ \hat{g}^{(3)}_{a+1},\hat{g}^{(3)}_{a+4} \right]  & = & -\hat{g}^{(3)}_{a+7} \hspace{0.4cm} \label{sl14} \\
\left[ \hat{g}^{(3)}_{a+1},\hat{g}^{(3)}_{a+5} \right]  & = &  0, \hspace{0.4cm}  
 \hspace{0.4cm} a=1 \dots 8  \hspace{0.5cm} {\rm mod} \; 8,1 .  \hspace{1.0cm}   \label{sl15}
\end{eqnarray}
with just one non-zero structure constant on the RHS in each of Eqs.~\ref{sl12}-\ref{sl14}. 
We give here explicitly the (complex) transformation 
of Eqs.~\ref{sl12}-\ref{sl15} into the Chevalley basis: 
\begin{equation}
\left( \begin{matrix} \hat{g}'_1 \cr \hat{g}'_2 \cr \hat{g}'_3 \cr \hat{g}'_4 
                    \cr \hat{g}'_5 \cr \hat{g}'_6 \cr \hat{g}'_7 \cr \hat{g}'_8 \end{matrix} \right) =
\left( \begin{matrix}
1 & 0 & 0 & 0 & -1 & 0 & 0 & 0 \cr 
\omega & 0 & 0 & 0 & -\bar{\omega} & 0 & 0 & 0 \cr 
0 & 0 & 0 & \frac{-i}{\sqrt{3}} & 0 & \frac{-i}{\sqrt{3}} & \frac{i}{\sqrt{3}} & 0\cr 
0 & \frac{-i\bar{\omega}}{\sqrt{3}} & \frac{i\omega}{\sqrt{3}} & 0 & 0 & 0 & 0 & \frac{-i}{\sqrt{3}} \cr
0 & 0 & 0 & \frac{-i}{\sqrt{3}} & 0 & \frac{-i\bar{\omega}}{\sqrt{3}} & \frac{i\omega}{\sqrt{3}} & 0 \cr
0 & \frac{-i}{\sqrt{3}} &  \frac{i}{\sqrt{3}} & 0 & 0 & 0 & 0 & \frac{-i}{\sqrt{3}} \cr
0 & 0 & 0 & \frac{-i}{\sqrt{3}} & 0 &  \frac{-i\omega}{\sqrt{3}} & \frac{i\bar{\omega}}{\sqrt{3}}  & 0 \cr
 0 & \frac{-i\omega}{\sqrt{3}} & \frac{i\bar{\omega}}{\sqrt{3}} & 0 & 0 & 0 & 0 & \frac{-i}{\sqrt{3}} \end{matrix} \right)
\left( \begin{matrix} \hat{g}^{(3)}_1 \cr \hat{g}^{(3)}_2 \cr \hat{g}^{(3)}_3 \cr \hat{g}^{(3)}_4 \cr 
                 \hat{g}^{(3)}_5 \cr \hat{g}^{(3)}_6 \cr \hat{g}^{(3)}_7 \cr \hat{g}^{(3)}_8 \end{matrix} \right),
\end{equation}
where $\omega=\exp(2\pi i/3)$ and $\bar{\omega}=\exp(-2\pi i/3)$ are the two complex cube roots of unity.
The resulting (real) algebra, having metric signature $(-^3,+^5)$,  
is sl(3,R): 
\begin{eqnarray} 
[\hat{g}'_1,\hat{g}'_2]=0 &
[\hat{g}'_1,\hat{g}'_3]=2\hat{g}'_3 &
[\hat{g}'_1,\hat{g}'_4]=\hat{g}'_4  \hspace{0.3cm}
[\hat{g}'_1,\hat{g}'_5]=-\hat{g}'_5   \nonumber \\
& [\hat{g}'_1,\hat{g}'_6]=-2\hat{g}'_6 &
[\hat{g}'_1,\hat{g}'_7]=-\hat{g}'_7 \hspace{0.3cm} [\hat{g}'_1,\hat{g}'_8]=\hat{g}'_8   \nonumber \\
& [\hat{g}'_2,\hat{g}'_3]=-\hat{g}'_3  &
[\hat{g}'_2,\hat{g}'_4]=\hat{g}'_4  \hspace{0.3cm} [\hat{g}'_2,\hat{g}'_5]=2\hat{g}'_5  \nonumber \\
& [\hat{g}'_2,\hat{g}'_6]=\hat{g}'_6  &
[\hat{g}'_2,\hat{g}'_7]=-\hat{g}'_7 \hspace{0.3cm}  [\hat{g}'_2,\hat{g}'_8]=-2\hat{g}'_8  \nonumber \\
& [\hat{g}'_3,\hat{g}'_5]=-\hat{g}'_4 &
[\hat{g}'_3,\hat{g}'_6]=\hat{g}'_1  \hspace{0.3cm} [\hat{g}'_3,\hat{g}'_7]=\hat{g}'_8  \nonumber \\
 & [\hat{g}'_4,\hat{g}'_6]=\hat{g}'_5  &
[\hat{g}'_4,\hat{g}'_7]=\hat{g}'_1+\hat{g}'_2  \hspace{0.3cm} [\hat{g}'_4,\hat{g}'_8]=-\hat{g}'_3   \nonumber \\
& & [\hat{g}'_5,\hat{g}'_7]=-\hat{g}'_6 \hspace{0.3cm} [\hat{g}'_5,\hat{g}'_8]=\hat{g}'_2  \nonumber \\
& & [\hat{g}'_6,\hat{g}'_8]=\hat{g}'_7 \label{sl3ry}
\end{eqnarray} 
as is immediately evident 
switching to the more familiar notation:
\mbox{$\hat{g}'_1=H^{\alpha}$, $\hat{g}'_2=H^{\beta}$,}
$\hat{g}'_3=E^{\alpha}$, $\hat{g}'_4=E^{{\alpha +\beta}}$, $\hat{g}'_5=E^{\beta}$,
$\hat{g}'_6=E^{-\alpha}$, $\hat{g}'_7=E^{-(\alpha+\beta)}$, $\hat{g}'_8=E^{-\beta}$ 
(see e.g. Ref.~\cite{[{}]FUCHS1,*FUCHS2}).
The corresponding inverse transformation
from the Chevalley basis (Eq.~\ref{sl3ry}) back
to the cyclic basis for su(2,1) (Eqs.~\ref{sl12}-\ref{sl15}) is then:
\begin{equation}
\left( \begin{matrix} \hat{g}^{(3)}_1 \cr \hat{g}^{(3)}_2 \cr \hat{g}^{(3)}_3 \cr \hat{g}^{(3)}_4 \cr 
                                  \hat{g}^{(3)}_5 \cr \hat{g}^{(3)}_6 \cr \hat{g}^{(3)}_7 \cr \hat{g}^{(3)}_8 \end{matrix}  \right) =
\left( \begin{matrix}
\frac{i\bar{\omega}}{\sqrt{3}} & \frac{-i}{\sqrt{3}}
                  & 0 & 0 & 0 & 0 & 0 & 0 \cr 
 0 & 0 & 0 & \frac{i\omega}{\sqrt{3}} & 0 
                & \frac{i}{\sqrt{3}} & 0 & \frac{i\bar{\omega}}{\sqrt{3}} \cr 
0 & 0 & 0 & \frac{-i\bar{\omega}}{\sqrt{3}} & 0 
                & \frac{-i}{\sqrt{3}} & 0 & \frac{-i\omega}{\sqrt{3}} \cr 
 0 & 0 & \frac{i}{\sqrt{3}} & 0 & \frac{i}{\sqrt{3}} & 0 & \frac{i}{\sqrt{3}} & 0 \cr
 \frac{i\omega}{\sqrt{3}} & \frac{-i}{\sqrt{3}}
                  & 0 & 0 & 0 & 0 & 0 & 0 \cr
 0 & 0 & \frac{i}{\sqrt{3}} & 0 &  \frac{i\omega}{\sqrt{3}}  
                                  & 0 &  \frac{i\bar{\omega}}{\sqrt{3}}  & 0 \cr
 0 & 0 & \frac{-i}{\sqrt{3}} & 0 &  \frac{-i\bar{\omega}}{\sqrt{3}}  
                                  & 0 &  \frac{-i\omega}{\sqrt{3}}  & 0 \cr
 0 & 0 & 0 & \frac{i}{\sqrt{3}} & 0 & \frac{i}{\sqrt{3}} & 0 & \frac{i}{\sqrt{3}} \end{matrix} \right)
\left( \begin{matrix} \hat{g}'_1 \cr \hat{g}'_2 \cr \hat{g}'_3 \cr \hat{g}'_4 
                            \cr \hat{g}'_5 \cr \hat{g}'_6 \cr \hat{g}'_7 \cr \hat{g}'_8 \end{matrix} \right).
\end{equation}

The (complex) transformation $C^{(13)}$ from su(2,1) (Eqs.~\ref{sl12}-\ref{sl15}) to su(3) (Eqs.~\ref{su12c2}-\ref{su15c2}) is given by:
\begin{equation}
C^{(13)}={\rm circ}\{(1+i),0,0,0,(1-i),0,0,0\}
\end{equation}
and the corresponding inverse transformation is given by:
\begin{equation}
C^{(31)}=(C^{(21)})^{-1}={\rm circ}\{(1-i)/4,0,0,0,(1+i)/4,0,0,0\} .
\end{equation}

Finally, we note, with admittedly some benefit of hindsight,
that our cyclic form Eqs.~\ref{su12c2}-\ref{su15c2} for su(2,1) 
might perhaps have been derived more analytically,
possibly without the need to resort to a computer search, 
exploiting the
`trigonometric structure constant' (TSC) concept \cite{[{}]PATER,*FAIR1,*FAIR2,*FAIR3},
by adopting suitable re-phasings, re-scalings and re-orderings 
of the TSC operators in the A$_2$ case.
Consequently,
recalling our (tentative)  
parting conjecture of the main text above,
an opportunity arises 
to try to exploit the A$_4$ TSCs 
to work towards a cyclic basis (if such exists)
for the real algebra su(5).
Then, with all 24 grand-unified 
vector bosons of the (unbroken) SU(5) theory  \cite{GUTS}
appearing on an equal footing from the outset,
we would finally have our `manifest bosonic democracy'.

As a possible step in this direction,
we present here a `bi-cyclic' form which we have obtained
for the real algebra su(5) itself,
i.e.\ having metric signature $(-^{24})$. 
We refer to this form as `bi-cyclic' because
coefficient sets alternate as we step the indices by one,
such that 
the `odd' brackets below are cyclic in themselves
as are the corresponding `even' ones
(which in fact differ from their `odd' counterparts
only by a factor of the golden ratio $\eta=(1+\sqrt{5})/2$
and in some cases by some re-arrangement of signs).
Given this `bi-cyclicity' and the total anti-symmetry inherent here,
the following brackets are sufficient to fix all non-zero structure constants:
\begin{eqnarray}
\left[\hat{g}^{su(5)}_{a+1},\hat{g}^{su(5)}_{a+3}\right] & = & 
(-\hat{g}^{su(5)}_{a+4}+\hat{g}^{su(5)}_{a+6}
                                          -\hat{g}^{su(5)}_{a+16}-\hat{g}^{su(5)}_{a+18})  \hspace{4mm} \nonumber \\
\left[\hat{g}^{su(5)}_{a+2},\hat{g}^{su(5)}_{a+4}\right] & = & 
\eta (-\hat{g}^{su(5)}_{a+5}+\hat{g}^{su(5)}_{a+7}
                                          -\hat{g}^{su(5)}_{a+17}-\hat{g}^{su(5)}_{a+19})  \hspace{4mm} \nonumber \\
  \,             &         &       \nonumber \\ 
\left[\hat{g}^{su(5)}_{a+1},\hat{g}^{su(5)}_{a+9}\right] & = & 
\eta (-\hat{g}^{su(5)}_{a+5}-\hat{g}^{su(5)}_{a+8}
                                          +\hat{g}^{su(5)}_{a+17}-\hat{g}^{su(5)}_{a+20})  \hspace{4mm} \nonumber \\
\left[\hat{g}^{su(5)}_{a+2},\hat{g}^{su(5)}_{a+10}\right] & = & 
(-\hat{g}^{su(5)}_{a+6}+\hat{g}^{su(5)}_{a+9}
                                          +\hat{g}^{su(5)}_{a+18}+\hat{g}^{su(5)}_{a+21})  \hspace{4mm} \nonumber \\
  \,             &         &       \nonumber \\ 
\left[\hat{g}^{su(5)}_{a+1},\hat{g}^{su(5)}_{a+5}\right] & = & 
\eta (+\hat{g}^{su(5)}_{a+9}-\hat{g}^{su(5)}_{a+12}
                                          +\hat{g}^{su(5)}_{a+21}+\hat{g}^{su(5)}_{a+24})  \hspace{4mm} \nonumber \\
\left[\hat{g}^{su(5)}_{a+2},\hat{g}^{su(5)}_{a+6}\right] & = & 
(+\hat{g}^{su(5)}_{a+10}+\hat{g}^{su(5)}_{a+13}
                                          +\hat{g}^{su(5)}_{a+22}-\hat{g}^{su(5)}_{a+1})  \hspace{4mm} \nonumber \\  
  \,             &         &       \nonumber \\ 
\left[\hat{g}^{su(5)}_{a+1},\hat{g}^{su(5)}_{a+11}\right] & = & 
(-\hat{g}^{su(5)}_{a+2}+\hat{g}^{su(5)}_{a+4}
                                          +\hat{g}^{su(5)}_{a+14}+\hat{g}^{su(5)}_{a+16})  \hspace{4mm} \nonumber \\
\left[\hat{g}^{su(5)}_{a+2},\hat{g}^{su(5)}_{a+12}\right] & = & 
\eta (-\hat{g}^{su(5)}_{a+3}+\hat{g}^{su(5)}_{a+5}
                                          +\hat{g}^{su(5)}_{a+15}+\hat{g}^{su(5)}_{a+17})  \hspace{4mm} \nonumber \\
 & & \hspace{1.0cm}  a=2, 4,\dots24  \hspace{0.3cm} {\rm mod} \; 24,1   
\label{bicyclic} 
\end{eqnarray}
 i.e. all structure constants are magnitude 1 or $\eta$ in modulus.
Note that elements 
separated by one quarter of a full cycle on the associated 24-gon,
i.e.\ by an index-count of 6 (mod 24), 
commute with each other:  
\begin{eqnarray}
\left[ \hat{g}^{su(5)}_{a+1},\hat{g}^{su(5)}_{a+7} \right]  & = &  
\left[ \hat{g}^{su(5)}_{a+1},\hat{g}^{su(5)}_{a+13} \right]  = 
\left[ \hat{g}^{su(5)}_{a+1},\hat{g}^{su(5)}_{a+19} \right] =0, \hspace{0.2cm} \nonumber \\
 & & \hspace{1.5cm}  a=1 \dots 24  \hspace{0.3cm} {\rm mod} \; 24,1   \hspace{0.4cm} 
\label{bicyclic2}
\end{eqnarray}
while the Killing form is diagonal ($\kappa^{su(5)}_{a b}
=-40\sqrt{5} \, \eta \, \delta_{ab}$).
Possibly this `bi-cyclic' form (Eqs.~\ref{bicyclic}-\ref{bicyclic2}) 
could be put into a fully cyclic form, e.g.\  by applying
suitable separate circulant transformations on even and odd generators.

\end{document}